\documentclass{article}
\usepackage{graphicx}
\usepackage{amsmath}
\usepackage{float}
\usepackage{subcaption}
\usepackage[T1]{fontenc}
\usepackage[utf8]{inputenc}
\usepackage{tgadventor}
\usepackage{txfonts}
\usepackage[T1]{fontenc}
\usepackage[T1]{fontenc}
\usepackage{ragged2e}
\usepackage[a4paper, left= 1.8cm, right= 1.8cm, top=1.5cm, textwidth=15.5cm]{geometry}
\usepackage{caption}
\usepackage{diagbox}
\usepackage{comment}
\usepackage{multicol,lipsum}
\usepackage[giveninits=true, backend=biber, style=numeric, sorting=none,maxbibnames=99]{biblatex} 

\usepackage[colorlinks]{hyperref}
\usepackage{blindtext}

\newenvironment{Figure}
  {\par\medskip\noindent\minipage{\linewidth}}
  {\endminipage\par\medskip}

\usepackage[export]{adjustbox}
\hypersetup{
     colorlinks = true,
     linkcolor = blue,
     anchorcolor = blue,
     citecolor = blue,
     filecolor = blue,
     urlcolor = blue
     }
\newcommand*\diff{\mathop{}\!\mathrm{d}}
\usepackage{todonotes}
\usepackage{graphicx}

\title{Studying Macro- and Mesoscopic Wetting Dynamics of a Spreading Oil Droplet Using Multiple Wavelength Interferometry} 

\bibliography{bib}

\author{Timo Richter\thanks{Institute for Fluid Mechanics and Aerodynamics, Department of Mechanical Engineering, TU Darmstadt, Germany}, Mathis Fricke\thanks{Institute for Mathematical Modelling and Analysis, Department of Mathematics, TU Darmstadt, Germany; ORCID: 0000-0002-6281-6617 }, Peter Stephan\thanks{Institute for Technical Thermodynamics Department of Mechanical Engineering, TU Darmstadt, Germany}, Cameron Tropea\footnotemark[1] \;and Jeanette Hussong\footnotemark[1]\;\thanks{E-mail address corresponding author: hussong@sla.tu-darmstadt.de}}

\begin{document}

\maketitle
\textbf{ABSTRACT:} In this study we present an interferometric technique based on multiple wavelengths to capture the transient free surface contour of nanoliter drops spreading on a wettable surface, in particular close to the three-phase contact line. Various data analysis procedures  are evaluated in terms of error and noise sensitivity. The technique allows an unambiguous determination of the local liquid film thickness for optical path differences up to $\Delta s \approx 3.19\,\mathrm{\mu m}$ without the need of a known reference height. Film thicknesses as low as $0.1\,\mathrm{\mu m}$ can be measured with the present optical configuration. 
The entire three-dimensional droplet shape is investigated for different capillary numbers, allowing also reliable measurements of the time-resolved contact angle.

\begin{multicols}{2}

\section{Introduction}
Thin film interferometry is a powerful tool to accurately determine liquid film thicknesses in a non-invasive manner, usually realized using light of a single wavelength, $\lambda$. In such cases each interference fringe corresponds to a given film thickness; however, a reference data point somewhere in the field of view is required to obtain an absolute thickness. Such a reference point/phase may not always be available. This difficulty is overcome in the present study using an interferometer with multiple wavelengths. By using multiple wavelengths, the absolute phase difference can be determined without a reference point, i.e. the $2\pi$ ambiguity can be resolved. The interferometer is demonstrated by measuring the spreading behaviour of an oil droplet on a highly wettable, sapphire surface.

Various types of multi-wavelength interferometers have been developed in the past \cite{interfero20162,yang2018review,van2012direct,desse2008digital}, exploiting the fact that the individual wavelengths create fringe patterns completely independent of one another. 
The existence of multiple fringe patterns with fringe spacing corresponding to different film thicknesses allows the absolute optical path difference to be determined. This necessitates using a colour sensitive camera, typically RGB, i.e. red, green and blue.
As a light source either a broadband light source combined with band-pass filters to narrow down the spectrum, or separate lasers of different wavelengths can be used. While experimental setups using a broadband light source are less expensive, the data evaluation  becomes more complicated due to a large colour channel overlap \cite{butler2016local}. Also, due to the low coherence length of broadband light sources, the interference patterns become less distinct  when the optical path length difference between the two interfering beams reaches values that are much higher than the wavelength of the light source. \\

While the spreading of larger spherical droplets with finite contact angle on smooth homogeneous surfaces is a well studied system \cite{tanner1979spreading, wang2007spreading,chenwada,starov2007wetting}, the spreading dynamics of nanoliter sized, perfectly wetting droplets is a subject of ongoing research. 
In the present study, the  shape of a nanoliter sized droplet is measured during its spreading on a surface by means of colour interferometry and compared to existing theories in the literature.
While standard visualisation techniques such as brightfield or darkfield imaging  provide information on the macroscopic droplet shape, more  techniques with a much higher resolution, such as total internal reflection spectroscopy and ellipsometry are limited to extremely small film heights \cite{franken2014dynamics,beaglehole1989profiles}.
Colour interferometry complements these measurement techniques, as it  covers a film thickness range lying between these ranges  down to  mesoscopic scales. As there seems to be no consensus in the literature on the definition of the dimension of the mesoscopic region, we will refer to the region in which the droplet contour deviates from a spherical cap as mesoscopic.\\
Usually, interferometric measurements of droplet spreading are performed with droplet volumes of a few microliters, where the focus is set on measuring the contact line region \cite{chenwada,kavehpour2003microscopic}. A very early and seminal work was conducted by Chen and Wada \cite{chenwada}, who studied the shape of a spreading silicone oil drop in the contact line region using interferometry measurements. These experiments were carried out with droplet volumes ($V\approx 0.5\, \mu\text{l}$) on soda-lime glass plate. Reducing Eq.~\eqref{DGL} (introduced in section~\ref{sec:3}) into  dimensionless form, the authors showed that the advancing meniscus shapes collapsed onto a single shape. \\
However, in these studies only a small portion of the droplet contour was investigated. The influence of the mesoscopic regime on the overall wetting process of nanolitre droplets is not yet fully understood; in particular the contact angle is of great interest for modeling and simulation efforts, for instance using Volume of Fluid (VOF) methods. 
\section{Measurement principle}

\subsection{Thin film interference}
\begin{figure}[H]
	\centering
\includegraphics[width=0.45\textwidth]{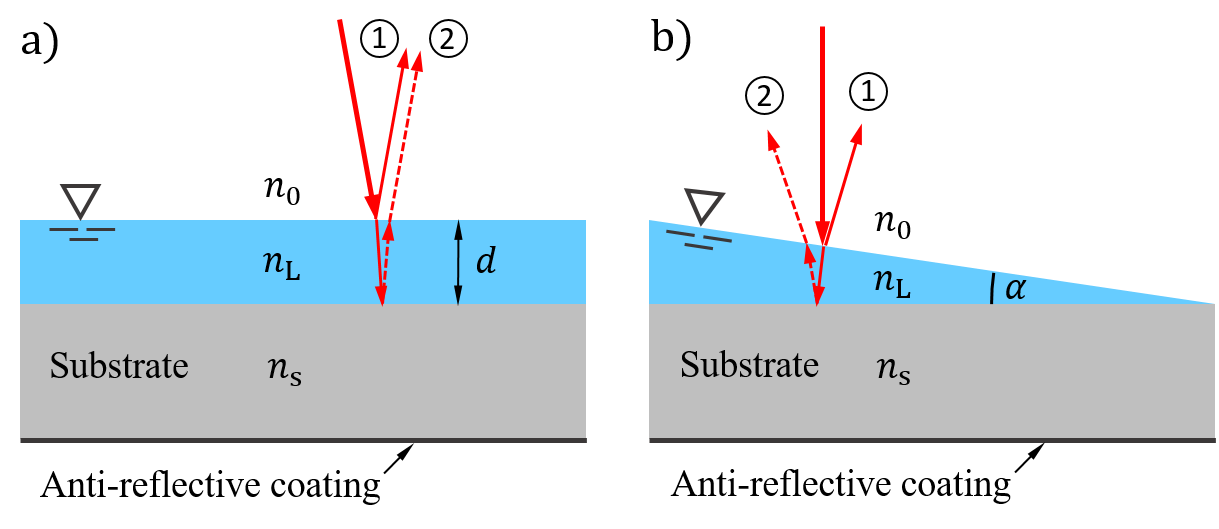}
	\caption{a) Light ray oriented obliquely to the substrate and impinging obliquely onto a liquid film surface of constant thickness. b) Light ray oriented normal to the substrate but impinging obliquely onto a liquid film of varying thickness.}
	\label{fig:ZeichnungStrahlen}
\end{figure}

The principle of thin film interferometry is illustrated in Figs~\ref{fig:ZeichnungStrahlen}a) and b). A light ray impinges on the free surface of a liquid film of refractive index $n_L$. According to the Fresnel equations, part of the light is reflected (Ray 1) and part refracted. The refracted part then traverses the film of thickness, before a portion is then again reflected at the substrate surface with refractive index $n_S$. This reflected portion then exits the free surface through refraction (Ray 2) and interferes with the initial reflected ray, resulting in interference, described by \cite{butler2016local,hariharan1992basics}
\begin{equation}
I=I_{1}+I_{2}+2\sqrt{I_{1}I_{2}}\cdot \cos (\Delta \Phi)
\end{equation}
where, $I_1$ and $I_2$ are the intensities of the two interfering  waves  and $\Delta \Phi = 4\pi n_L d/\lambda + \Delta\varphi$ is the phase difference between them.  
Hence, the local film thickness $d$ is directly related to the phase shift  $\Delta \Phi$, which for every $2\pi$ results in an additional interference fringe. The optical path difference is given by $\Delta s = 2n_Ld$.  Here, $\Delta\varphi$ denotes the phase offset that occurs when a phase change is induced at the reflecting interfaces. With the present configuration there is a phase offset of $\Delta\varphi = \pi$ at each of the reflecting surfaces of rays 1 and 2, eliminating this term from the $\Delta \Phi$ expression. Thus the phase difference can be expressed as
\begin{equation}
\Delta \Phi = 2\pi\Delta s/\lambda
\label{eq:deltaPhi}
\end{equation}

Depending on the refractive index of the substrate, $n_S$, the incident refracted ray may also be refracted into the substrate, only a portion being reflected at the substrate upper surface. This is indeed the case using the sapphire glass substrate in the present study. For this reason, an anti-reflection coating was applied to the lower surface of the sapphire substrate  to reduce the intensity of any   rays from the lower surface being reflected back into the interferometer (which are therefore not shown in the figure).
The Fresnel equations can be strongly simplified for the case of perpendicular or a small angles of incidence, as in the present case, and moreover, there remains no dependence on  polarisation. The reflection coefficient $R$ at the upper film surface is then given by
\begin{equation}
R = \Biggl(      \frac{n_0-n_L}{n_0+n_L}  \Biggr)^2,
\end{equation}
Preserving total energy, the  transmission coefficient $T$ can be obtained as:
\begin{equation}
T = 1-R
\end{equation}
The same equations, using the corresponding refractive indices of the liquid and substrate, can be applied at the liquid-substrate surface to determine the intensity of the reflected and refracted rays at that interface. Using the approximate values of $n_0=1.00$, $n_L=1.403$ and $n_S=1.780$ yields $R=0.028 $ and $R=0.014$ at the two interfaces respectively, indicating that a large portion of the light is refracted into the liquid/substrate.
The fringe contrast or modulation $V$ is given by:
\begin{equation}
V = \frac{2\sqrt{I_1I_2}}{I_1+I_2}
\end{equation}
A maximum fringe contrast of $V=1$ is obtained for $I_1=I_2$, corresponding to total constructive and destructive interference. This value of $V$ will depend largely on the respective refractive indices involved.  Furthermore,  the liquid should be highly non-absorbing at the wavelengths used for the interferometer. \\
\subsection{Thin film interference for spatially varying film heights}
The case is now examined in which the liquid surface is at an angle $\alpha$ with respect to the substrate, as pictured in  Fig.~\ref{fig:ZeichnungStrahlen}b). In this situation the two rays 1 and 2 are divergent above the film and any imaging optics above the film must ensure a numerical aperture large enough to capture both rays. The two rays now intersect just beneath the film surface; hence, the interference fringes captured in an image are the virtual fringes inside the film \cite{hariharan1992basics}.
Alternatively, illumination and observation can be from below through a transparent substrate  \cite{van2012direct,butler2016local}, which can be an appropriate configuration for measuring for instance a vapour or air film beneath an impacting droplet \cite{van2012direct}. 
It becomes apparent that low free surface angles $\alpha$ are preferred for measurements since they simplify the imaging optics. In the same sense, large undulations of the free surface may not be captured and resolved by the interferometer.
\subsection{Multiple colour interferometry}
\begin{figure}[H]
	\centering
  \includegraphics[width=0.39\textwidth]{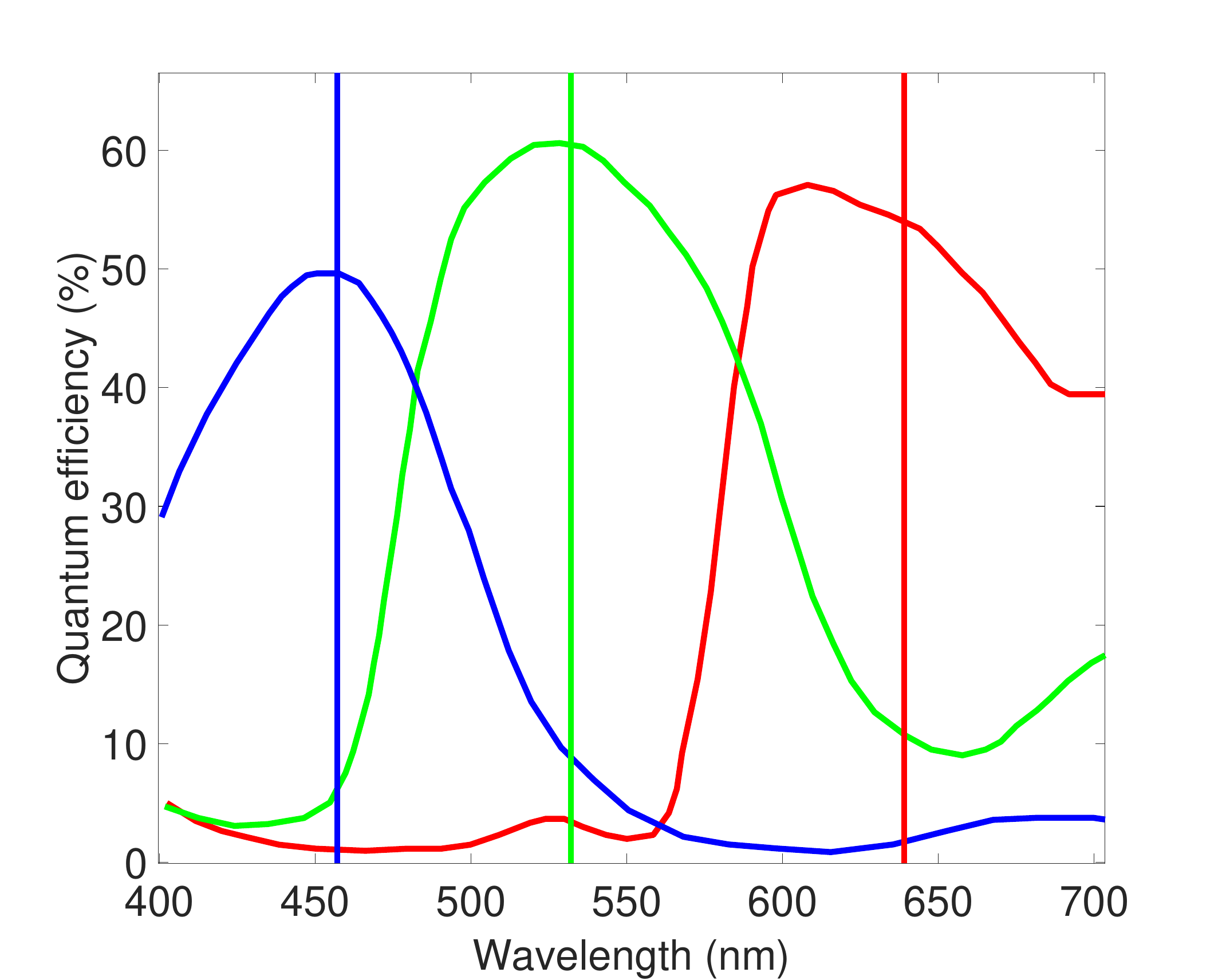}
	\caption{Spectral sensitivity of the colour CMOS camera and the selected laser wavelengths.}
	\label{fig:quantum}
\end{figure}

\begin{figure}[H]
	\centering
  \includegraphics[width=0.45\textwidth]{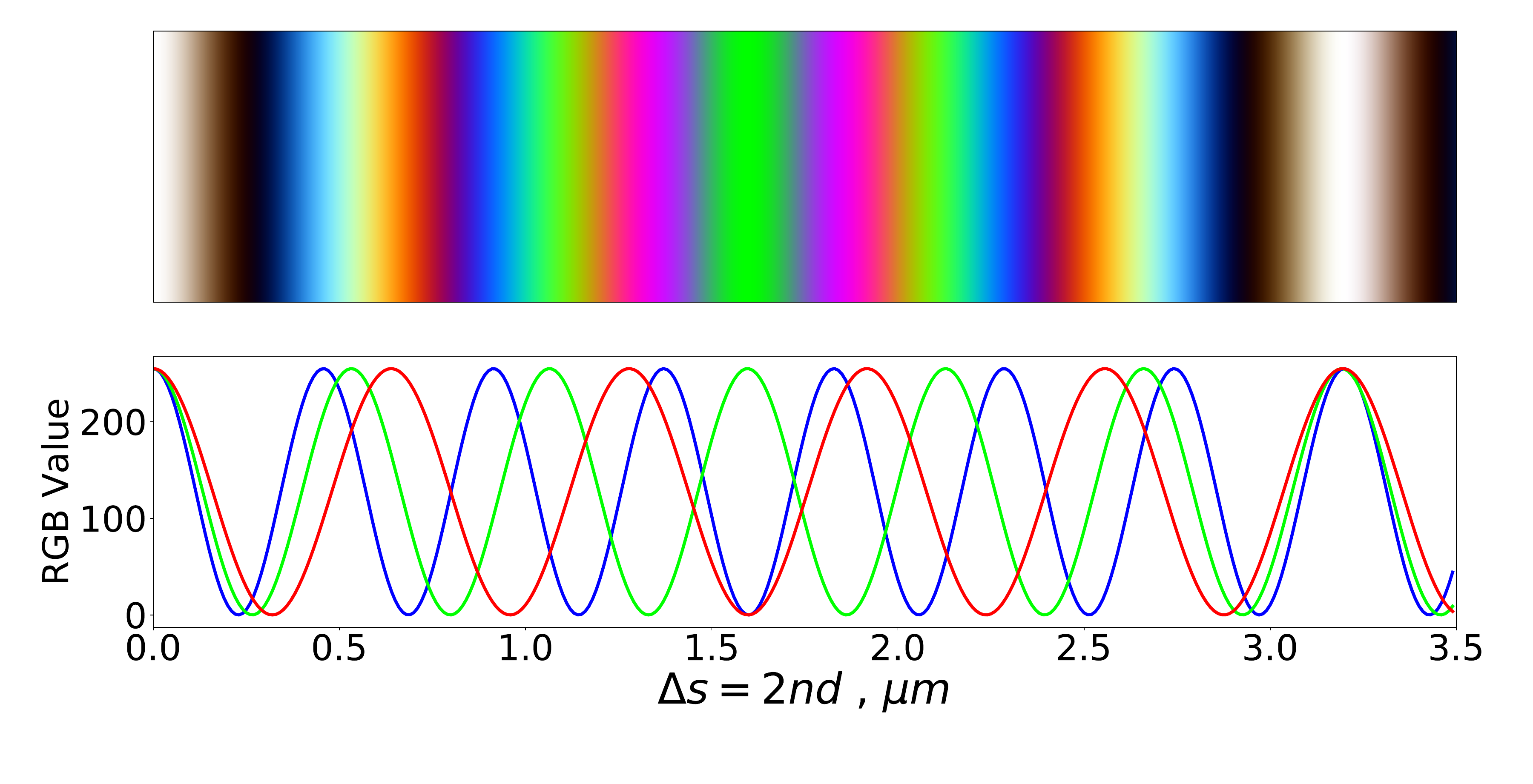}
	\caption{ Three laser wavelengths ($\lambda_B=457\,\text{nm}$, $\lambda_G=532\,\text{nm}$ and $\lambda_R=639\,\text{nm}$)  of equal intensity superimposed on one another.  The colour pattern repeats after $\Delta s = 6\lambda_R$.}
	\label{fig:farbpattern}
\end{figure}

Colour interferometry involves separate interferometers operating at multiple wavelengths simultaneously with no interference among the individual fringe patterns. This is particularly appropriate and necessary when measuring dynamic systems, such as the spreading oil droplet investigated in the present study, since the interferometers at different wavelengths cannot be invoked sequentially. To achieve this, three different laser sources with very narrow line widths were used, combined optically into a single illuminating beam. The wavelengths were selected to match the wavelengths at which the RGB CMOS camera exhibited its highest quantum efficiencies. This is illustrated in Fig.~\ref{fig:quantum}, in which the camera spectral sensitivities and the chosen laser wavelengths at $457\,\text{nm}$, $532\,\text{nm}$ and $639\,\text{nm}$ are shown.
A synthetic superposition of the three infinitely narrow wavelengths with 100\% fringe contrast (i.e. equal intensities or $V=1$) is pictured in Fig.~\ref{fig:farbpattern}. This simulation illustrates that a unique colour results up to an optical path difference of about $\Delta s \approx 1.6\,\mu\text{m} $. The pattern itself is repeated after $\Delta s \approx 3.2\,\mu\text{m} $ which corresponds to the maximum value of $\Delta s$, which can be uniquely determined. The colour information can be directly assigned to a film thickness without any reference points. It should be noted that this range could be further increased by selecting different wavelengths.
\section{Spreading of highly wetting liquids}
\label{sec:3}
The spreading properties of liquid drops on smooth solid surfaces has long been a topic of research, focusing primarily the spreading rate and the variation of dynamic contact angle \cite{tanner1979spreading,starov2007wetting,de1985wetting}. \\
For partial wetting, a finite contact angle  at the three-phase contact line region arises, which nominally is given by the well-known Young's equation:
\begin{equation}
    \cos(\theta_0)=\frac{\gamma_\mathrm{S}-\gamma_\mathrm{SL}}{\gamma_\mathrm{L}}
\label{Young}
\end{equation}
whereby  $\gamma_\mathrm{S}$, $\gamma_\mathrm{L}$ and $\gamma_\mathrm{SL}$ are the interfacial tensions of solid-gas, liquid-gas and solid-liquid respectively.\\
If the system is in thermodynamic equilibrium, i.e. $\gamma_\mathrm{S} = \gamma_\mathrm{SL} +\gamma_\mathrm{L}$, this represents complete wetting, where the equilibrium contact angle vanishes ($\theta_0=0$). Since in many experimental situations the system is not in perfect thermodynamic equilibrium, the right-hand side of Young's equation can actually exceed unity. The  spreading coefficient $S$ can be defined as follows \cite{bonn2009wetting}:
\begin{equation}
 S = \gamma_\mathrm{S} -\gamma_\mathrm{SL}-\gamma_\mathrm{L}
\label{Spreading coefficient}
\end{equation}
In the case of complete spreading, $S\geq 0$ holds.
Using this expression,  Young's equation becomes:
\begin{equation}
    S = \gamma_\mathrm{L} (\cos \theta_0 -1)
\end{equation}
From this equation it is apparent that in the case of a positive spreading coefficient,  Young's equation is not able to relate the spreading coefficient to an equilibrium contact angle. \\
While $S$ is often described as a measure of how rapidly a liquid can spread on a surface, the macroscopic wetting dynamics are expected to be independent of the exact value of the spreading coefficient \cite{de1985wetting}. \\
This was explained by  de Gennes \cite{de1985wetting} using the existence of a  precursor film, which  moves ahead of the contact line and whose existence has been proven experimentally for completely and also partially wetting fluids. The precursor film is created by surface forces that are relevant in the microscopic region near the liquid-solid interface \cite{beaglehole1989profiles,kavehpour2003microscopic,popescu2012precursor,franken2014dynamics}. In the case of non-polar liquids, only van der Waals forces should be relevant, while in general also electrostatic forces can be present. \\
While the precursor film has been the subject of intense investigation,  formulations exist for the macroscopic contact angle under the assumption that a macroscopic meniscus is moving with a constant shape, a low velocity, $U$, and that gravitational effects can be neglected. Using the lubrication approximation one finds \cite{bonn2009wetting}
\begin{equation}
 \frac{3\mu U}{h(x)^2}=-\frac{\diff}{\diff x}\biggl[ \gamma_L \frac{\diff^2 h(x)}{\diff x^2} + \frac{A}{6\pi h(x)^3}  \biggr],
\label{DGL}
\end{equation}
where $h(x)$ is the height of the droplet at the spatial coordinate $x$, $\mu$ is the viscosity, and $A$ is the Hamaker constant. This equation requires the assumption of small angles where $\diff h(x)/\diff x \ll 1$ holds and the curvature can be approximated by the second derivative of the film height. \\
The last term in Eq.~\eqref{DGL} describes the disjoining pressure in the case where only van der Waals forces are of importance. In the case where $A>0$ holds, the formation of a liquid wetting layer is favored, since this case corresponds to a repulsive force between the liquid-solid and liquid-vapor interface. Generally, van der Waals forces can be divided into dipolar interactions and induced dipolar interactions, while usually the latter dominate if the liquid is not highly polar. The corresponding Hamaker constant can then be calculated using the refractive indices of the substances \cite{israelachvili2011intermolecular}
\begin{equation}
 A = -\frac{3\pi\hbar\nu_{UV}}{8\sqrt{2(n_1^2+n_3^2)(n_2^2+n_3^2)}}\cdot \frac{(n_1^2-n_3^2)(n_2^2-n_3^2)}{\sqrt{n_1^2+n_3^2}+\sqrt{n_2^2+n_3^2}}
\label{hamaker}
\end{equation}
where $n_1$,$n_2$ and $n_3$ are the refractive indices of the solid, vapor and liquid phase, respectively. $\hbar$ is the reduced Planck constant and $\nu_{\text{UV}} \approx 2\cdot 10^{16}\,\text{Hz}$ is the UV frequency, where the dielectric permittivity approaches unity and the difference between the refractive indices of both materials vanishes \cite{bonn2009wetting,butt2018surface}. Usually it is assumed that the surface forces start to play a significant role if the liquid film thickness is smaller than $100\,\text{nm}$ \cite{de1985wetting}. \\
Equation \eqref{DGL} can be written in the following manner:
\begin{equation}
 3Ca/h^2=-h^{\prime\prime\prime}+3a^2h^{\prime}/h^4
\label{DGL2}
\end{equation}
Here, $Ca=\mu U/ \gamma_L$ is the capillary number and $a=\sqrt{A/6\pi \gamma_L}$ is a microscopic length scale that is usually expected to be of order $\approx 0.1\,\text{nm}$. In the case where one is only interested in the macroscopic flow properties, and under the assumption that the surface force term can be neglected, a  solution to this equation has been given by Voinov \cite{voinov1976hydrodynamics}. Assuming that the solution should be asymptotic to an outer solution $h_{out}(x)$ with vanishing curvature $h_{out}^{\prime\prime}(\infty)=0$ far away from the contact line, and setting the boundary condition that the contact angle approaches a finite value $\theta_{mic}$ on a microscopic length scale $l_{mic}$, one obtains the following solution:
\begin{equation}
\theta_{app}^3(x) \approx h^{\prime 3}(x) = \theta_{mic}^3+9Ca\ln(x/l_{mic})
\label{cox-Voinov}
\end{equation}
Equation \eqref{cox-Voinov} is usually known as the Cox-Voinov law. It says that in the case where the contact line is moving with a velocity $U\neq 0$, the apparent contact angle deviates from the contact angle on a microscopic scale, and that the apparent contact angle varies logarithmically with the distance from the contact line. \\
The third power of the apparent contact angle being proportional to the contact line speed $\theta_{app}^3\propto Ca$ has repeatedly been confirmed in different types of experiments such as capillary rise and droplet spreading \cite{wang2007spreading,hoffman1975study}. However, we want to emphasize that the boundary condition of the vanishing curvature in the outer region and the moving contact line with a constant meniscus shape might not be suitable assumptions in the case of very small spreading droplets.

\section{Experimental setup and procedure}
\begin{figure}[H]
	\centering
  \includegraphics[width=0.45\textwidth]{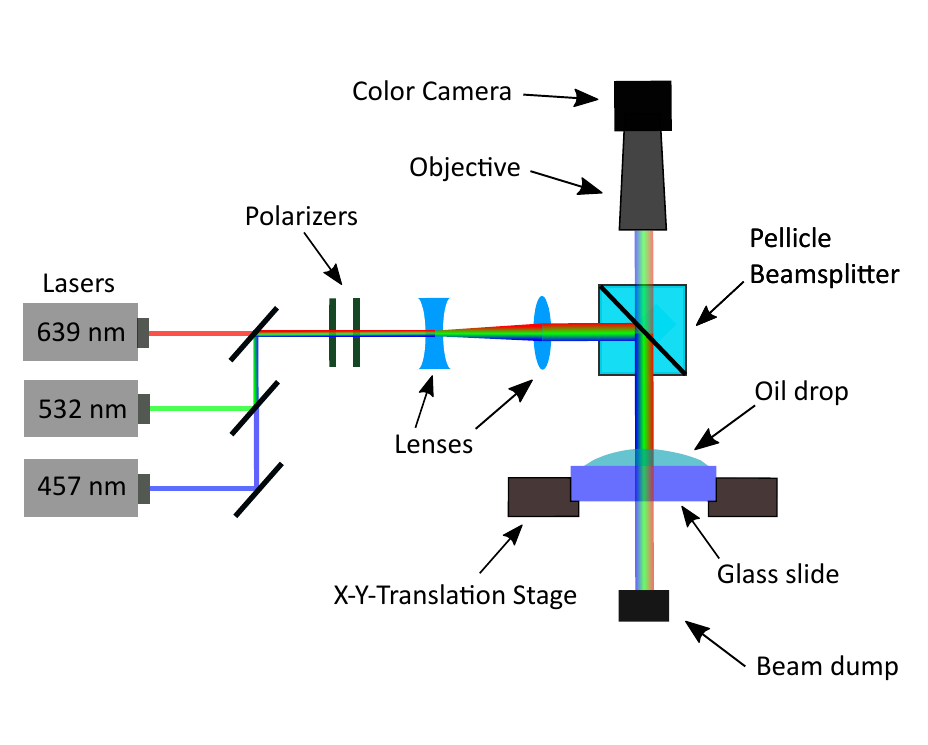}
	\caption{Sketch of the colour interferometer  used to investigate the thickness of the oil droplet. }
	\label{fig:versuchsaufbau}
\end{figure}

The experimental setup  used in this study is shown in Fig~\ref{fig:versuchsaufbau}. Beams from the three lasers are combined through a system of mirrors and then polarized using rotatable, linear polarizers. These polarizers are also used to equalize the light intensity of the three beams. This combined beam is then expanded to about 7mm using a convex-concave lens combination. A  pellicle beamsplitter  redirects the  beam toward the target sample, which is a drop of silicone oil placed onto a sapphire glass surface. The pellicle  type beam expander prevents ghost imaging, which is important when light with a high coherence length is used. In this way the incidence beam can be oriented vertically to the substrate as illustrated in Fig.~\ref{fig:ZeichnungStrahlen}b) without perturbing ghost images occurring in the recorded fringe pattern. The reflected light from the air-liquid and liquid-solid interface is then reflected back through the beamsplitter and into the colour camera (iDS UI-3080SE-C-HQ). The camera has a resolution of 2456 x 2054 pixels, a pixel size of $3.45\,\mu\text{m}$ and a frame rate of 27fps. A long distance microscope objective (Navitar) with a maximum magnification of 14 is used. A high precision X-Y translation stage is used to vary the field of view, while simultaneously serving as a calibration system for the spatial coordinates. \\
The substrate (sapphire glass, $n\approx 1.76$) was cleaned using acetone, ethanol and distilled water,  silicone oil (Sigma Aldrich, $\eta=4.565\,\text{mPas}$, $\gamma \approx 20\,\text{mN/m}$, $\rho = 0.913\,\text{g/ml}$, $n\approx 1,403$) was filtered  using a $2\,\mu\text{m}$ hydrophobic filter, and a small droplet of approximately $9\,\text{nl}$ was placed onto the smooth substrate using a syringe.

\section{Data analysis and image processing}
In the past, various approaches to processing and evaluating images from multiple colour interferometers have been implemented \cite{interfero20162,van2012direct,kitagawa2013thin}. Some use directly the  intensity information, which offers the possibility of a colour lookup table \cite{butler2016local} or the use of an alternative colour space \cite{van2012direct}. However these methods require a highly controlled light intensity of each wavelength throughout the entire field of view, which is often hard to realize in  experimental setups, when multiple light sources are used.  Furthermore, if the liquid exhibits different light adsorption coefficients at the different wavelengths, this necessitates a calibration step in the procedure. It is therefore preferable to make use of the phase information embedded in the interference patterns, which can be obtained using a Fourier data analysis. \\
The absolute phase $\Delta \Phi$ of each of the three interference patterns is directly connected to the optical path difference and therefore the local film thickness. The phase information is extracted from the image after first filtering the raw image in wavenumber space, i.e. first applying a two-dimensional Fourier transform, applying a band-pass filter, and then invoking an inverse Fourier transform to obtain an image with reduced white-noise content. The DC component is also removed,  since this value only contains the average intensity information. From the filtered image, the phase information can then be obtained using the real and imaginary parts of the Fourier filtered image $z(i,j)$:
\begin{equation}
\phi(i,j) = \mathrm{arctan2} \bigl(  \Im [z(i,j)] , \Re [z(i,j)]  \bigr)\, ; \, \phi \in [-\pi,+\pi]
\label{eq:atan2phase}
\end{equation}
 This provides the 'wrapped' phase angle $\phi(i,j)$ over the entire image.  However, to reconstruct the absolute height of the liquid, the unwrapped phase $\Phi(i,j)$ is required, since this is the quantity that is directly connected to $\Delta s$. In the case of multiple colour interferometry, the wrapped phase data can be obtained for all three wavelengths simultaneously, yielding a multitude of possible liquid film heights. However,  the  correct height of the liquid has to be identical in all three cases. Two approaches  to obtain the unwrapped phase information will now be discussed: the first using a lookup table; the second with a phase unwrapping algorithm. 

\subsection{Unwrapped phase  via a lookup table}
\label{subsec:lookup}
\begin{figure}[H]
	\centering
  \includegraphics[width=0.5\textwidth]{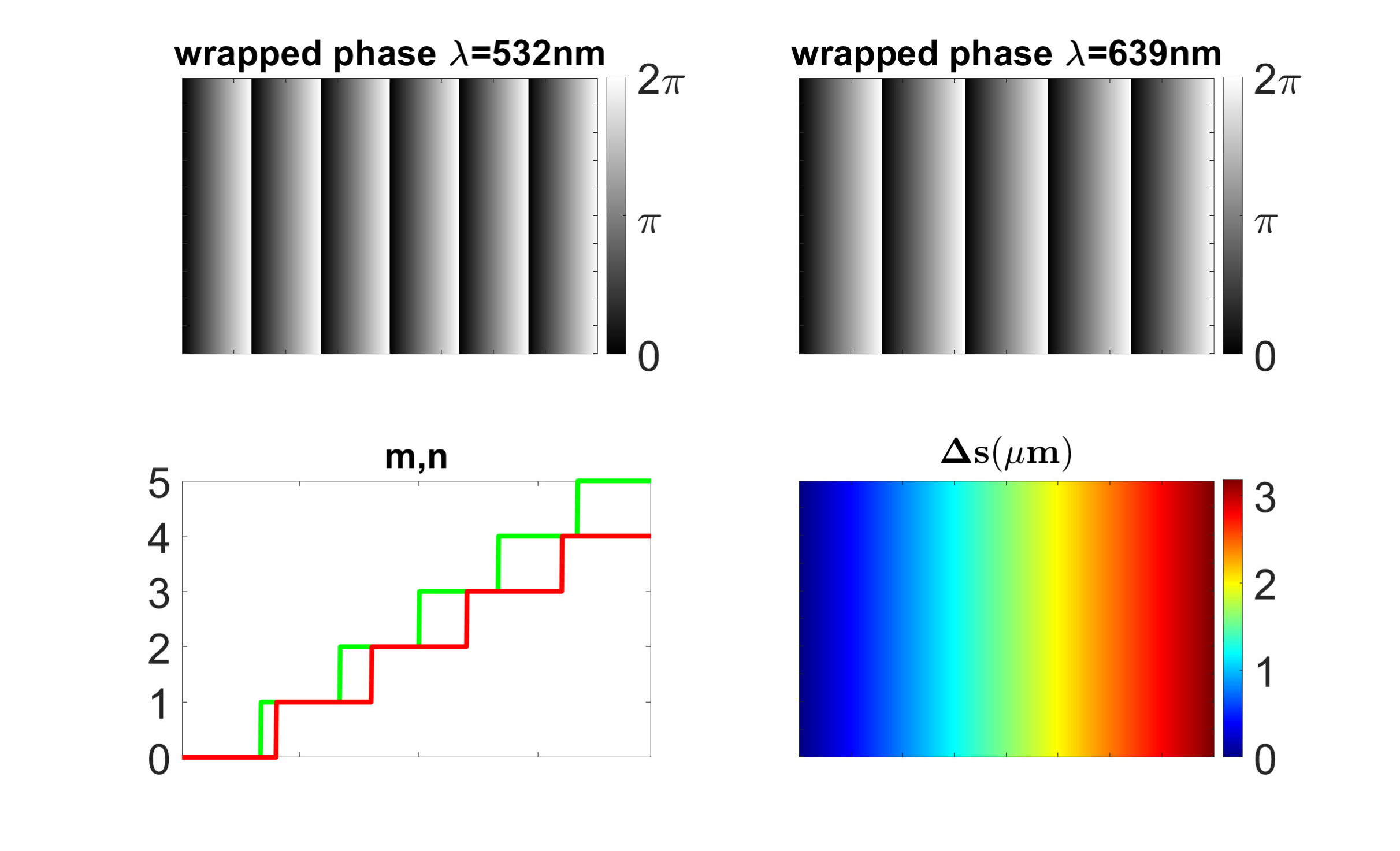}
	\caption{The wrapped phase data of two colour channels was used to extract the values of $m$ and $n$. Afterwards, these values can be used to calculate the unwrapped phase data, which is directly connected to $\Delta s$.}
	\label{fig:method01}
\end{figure}

The lookup table approach builds on the fact that the optical path difference $\Delta s$ is identical for the light of all implemented wavelengths \cite{long2014absolute}. This will be illustrated for the case of two wavelengths. The relationships between the unwrapped phases $\Phi_{\lambda_1}(i,j)$ and $\Phi_{\lambda_2}(i,j)$ at the two  wavelengths and the optical path difference $\Delta s$ are given by (see Eq.~(\ref{eq:deltaPhi}))
\begin{subequations}
\begin{align}
(2\pi / \lambda_1)\Delta s = \Phi_{\lambda_1} = 2\pi m+\phi_1 \\
(2\pi / \lambda_2)\Delta s = \Phi_{\lambda_2} = 2\pi n+\phi_2
\end{align}
\label{eq:gl}
\end{subequations}
whereby $m$ and $n$ are integers. Using these equations we can derive the  relationship:
\begin{align}
\frac{\lambda_1 \phi_1-\lambda_2 \phi_2}{2\pi} = n\lambda_2 - m\lambda_1 
\label{eq:Abgeleitet}
\end{align}
The left-hand side of Eq.~\eqref{eq:Abgeleitet} can be calculated from the experimental data and should be an integer value. The right-hand side contains the integers $m$ and $n$ that must be determined in order to reconstruct the unwrapped phase. First,  the interval of $\phi_i$ is shifted such that $\phi_i \in [0,2\pi]$ holds, which is achieved by adding $2\pi$ to all negative phase values. Then  the left-hand side of Eq.~\eqref{eq:Abgeleitet} is used to calculate an upper and lower bound:
\begin{align}
-\lambda_2< n\lambda_2 - m\lambda_1 <\lambda_1
\label{eq:Abgeleitet2}
\end{align}
The last  step  is to choose the interval of possible values for $m$ and $n$. It is evident that the lower bound of the interval is 0, while the upper boundary is the number of fringes that are visible in the image, if the reference fringe of height 0 is also contained. However, for the specific wavelengths that were chosen in our experimental setup, the following holds:
\begin{align}
 \frac{\lambda_R}{\lambda_G} \approx \frac{6}{5} \qquad  \frac{\lambda_R}{\lambda_B} \approx \frac{7}{5}
 \end{align}
This means that for optical path differences $\Delta s=2n_L d > 5 \lambda_R$ the phase pattern (and therefore the colour pattern) repeats itself, which can be seen in Fig.~\ref{fig:farbpattern}. A direct phase unwrapping using this method is therefore only possible for film thicknesses $ d < 5\lambda_R/2n$, which is about $1.14 \mathrm{\mu m}$ in the case of silicone oil. \\
This the limits the  interval range for $m$ and $n$. Using Eq.~\eqref{eq:Abgeleitet2}, we can then calculate a matrix containing the values of all possible combinations of pairs of $m$ and $n$ for the case where $\lambda_1=\lambda_R$ and $\lambda_2=\lambda_G$:
\begin{table}[H]
\begin{center}
\begin{tabular}{|l|cccccc|}
  \hline
  \diagbox{m}{n} & 0 & 1 & 2 & 3 & 4 & 5 \\
  \hline
  0      &    0    &     532     &            &       &     &   \\
  1      &        &     -107     &      425      &       &     &   \\
  2      &        &          &       -214     &    318   &     &   \\
  3      &        &          &            &    -321   &   211  &   \\
  4      &        &          &            &       &   -428  &  104 \\
 \hline
\end{tabular}
\caption{Table of all values in nanometers for $n\lambda_G - m\lambda_R$ that fulfill Eq.~\eqref{eq:Abgeleitet2}.   }
\label{fig:table}
\end{center}
\end{table} 
For each pixel in the obtained colour images, the left-hand side of Eq.~\eqref{eq:Abgeleitet} is calculated and assigned to its nearest value from the table. In this manner the values for $m$ and $n$ are obtained and can be used to calculate the unwrapped phase $\Phi$. An example is shown in Fig.~\ref{fig:method01}.
\subsection{Unwrapped phase via an algorithmic approach}
Numerous algorithms for phase unwrapping have been proposed in the past  \cite{long2014absolute,martinez2017fast}. Perhaps the most well-known is an iterative approach to improve the data quality and uses a concept from theoretical physics  \cite{martinez2017fast}. Although this approach is  appealing, simpler approaches are available, which are also noise insensitive and moreover, computationally  efficient \cite{long2014absolute}. The latter feature is particularly important in the present case, where entire, high resolution images must be processed. \\
In this second approach, the second spatial derivative of the phase angle $\phi(i,j)$ is estimated for each pixel value in the horizontal, vertical and both diagonal directions using all eight neighbouring pixel values. These four derivatives are denoted here as $D_1$, $D_2$, $D_3$ and $D_4$. Let us assume that the derivative $D_1$ refers to the second derivative of the phase angle $\phi(i,j)$ in horizontal direction, $D_1$ can be estimated using the difference equation:
\begin{align}
D_1(i,j)=\gamma[\phi(i-1,j)-\phi(i,j)]-\gamma[\phi(i,j)-\phi(i+1,j)]
\label{eq:seconddifference}
\end{align}
where $\gamma[\cdot]$ is an operator that removes steps of $2\pi$. The overall second derivative $D(i,j)$ of each pixel is calculated from all four derivative values:
\begin{align}
D(i,j) = \frac{1}{2}[D_1^2(i,j)+D_2^2(i,j)+D_3^2(i,j)+D_4^2(i,j)]^{1/2}
\label{eq:seconddifferenceG}
\end{align}
The reliability of a pixel can now be defined as the inverse of the second derivative:
\begin{align}
R(i,j)=\frac{1}{D(i,j)}
\label{eq:reliable}
\end{align}

The reliability value is  computed for each edge of every pixel  as the sum of the reliabilities for both pixels on that edge. The list of edges is sorted into an array from the highest to lowest reliability value. The two pixels that are connected via the most reliable edge are now sorted into one group. The two pixels that are connected through the second most reliable edge are sorted into another group, and so forth. If one of the two pixels of an edge is already part of one group, the new pixel is added to this group. If the difference between the new pixel and the one connected to it is larger than $\pi$, the value $2\pi$ is added to the new pixel. If the new pixel is also part of a group, each pixel in the smaller group is shifted by $2\pi$. This method is continued until every pixel is contained in the same group. \\
A major advantage of this procedure is  that experimental errors (i.e. damaged pixel areas) do not disturb the correct pixels, since they are sorted last. Nevertheless, the unwrapped phase data may still be falsely shifted by a multiple of $2\pi$, since the algorithm does not use a reference point, where the phase information is known. \\
If the unwrapped phase maps are obtained for at least two different wavelengths, we can  first obtain the unwrapped phase images. Then, both phase maps are shifted by multiples of $2\pi$ according to the ten possible combinations of $m$ and $n$ that were discussed in the framework of the lookup table discussed in section~\ref{subsec:lookup} (Table~\ref{fig:table}). The combination of $m$ and $n$ for which the difference between both maps is the smallest is then considered to be the correct one.  \\

\vspace {2cm}

\section{Results}
\begin{figure*}
	\centering
  \includegraphics[width=0.9\textwidth]{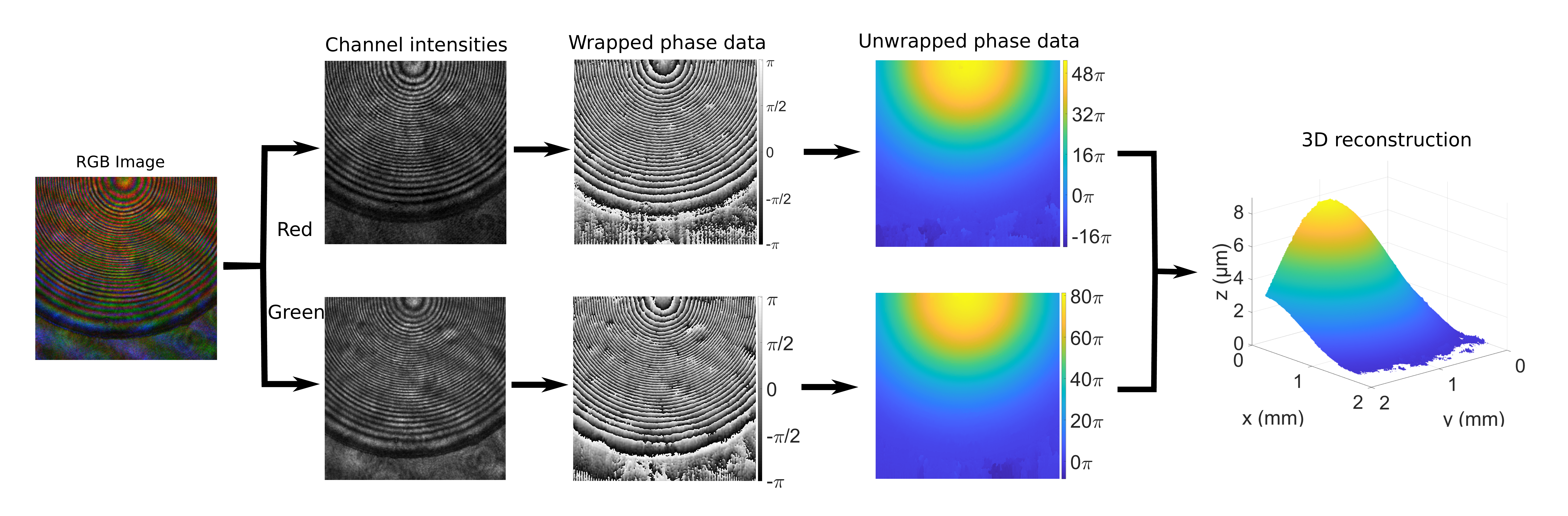}
	\caption{Data evaluation procedure for the  colour images. After reading out the red and green  channels, the wrapped phase data is obtained using Fourier analysis. Afterwards, the unwrapped phase maps are obtained using the phase unwrapping algorithm. Finally, the  combination of both maps is used to obtain a three-dimensional plot of the spreading silicone oil droplet.}
	\label{fig:method1}
\end{figure*}

Fig.~\ref{fig:method1} illustrates the evaluation process to derive the three-dimensional droplet contour from the RGB camera images. The evaluation process is demonstrated exemplary using the red and green colour channels. 
From both colour channels, the wrapped phase data was first obtained using the Fourier data analysis and applying Eq.~\eqref{eq:atan2phase}. Afterwards, the unwrapping algorithm was used for each wavelength. Using the fact that, according to Eq.~\eqref{eq:gl},  both phase maps must to correspond to the same film thickness,  an absolute phase map and therefore the absolute height distribution of the droplet can be calculated. In Fig.~\ref{fig:farbpattern} one distinct dark interference fringe is visible in the RGB image. Here, the optical path difference of about $300\,\text{nm}$ leads to destructive interference. In this region the film thickness equals approximately $100\,\text{nm}$. For even smaller film thicknesses, constructive interference leads to a maximum in intensity. However, the data analysis in this region is complicated. This is due to the fact that for constructive interference between the reflected light from the upper and lower liquid interfaces, roughly the same intensity is expected compared to the unwetted region, which can be calculated using the Fresnel reflection coefficients. Therefore, it is not possible to distinguish between the two cases using the Fourier analysis; hence, only liquid film thicknesses up to about $100\,\text{nm}$ can be evaluated. \\ 
Due to experimental uncertainties, the obtained height profile obtained from each of the colours is slightly different. The profile shown in Fig.~\ref{fig:method1} therefore corresponds to the averaged value from both  channels. We will interpret the magnitude of the difference between the two data sets as a measure for the accuracy of the data.  To illustrate this, Fig.~\ref{fig:errors}a) shows the height difference $|\Delta z|$ between both data sets. \\

\begin{Figure}
\begin{minipage}[b]{0.48\columnwidth}
\includegraphics[width=\textwidth]{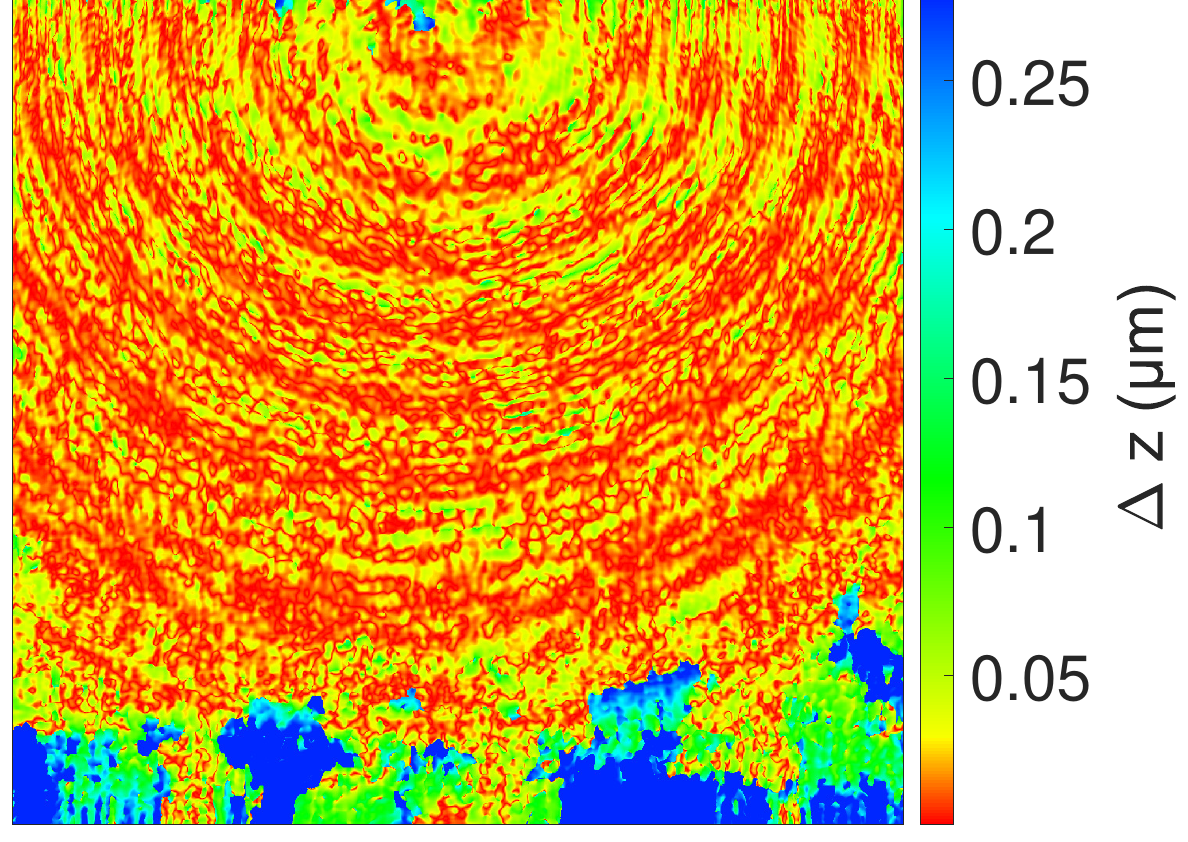}
\label{fig:errors1}
\end{minipage}
\hspace{0.1cm}
\begin{minipage}[b]{0.48\columnwidth}
\includegraphics[width=\textwidth]{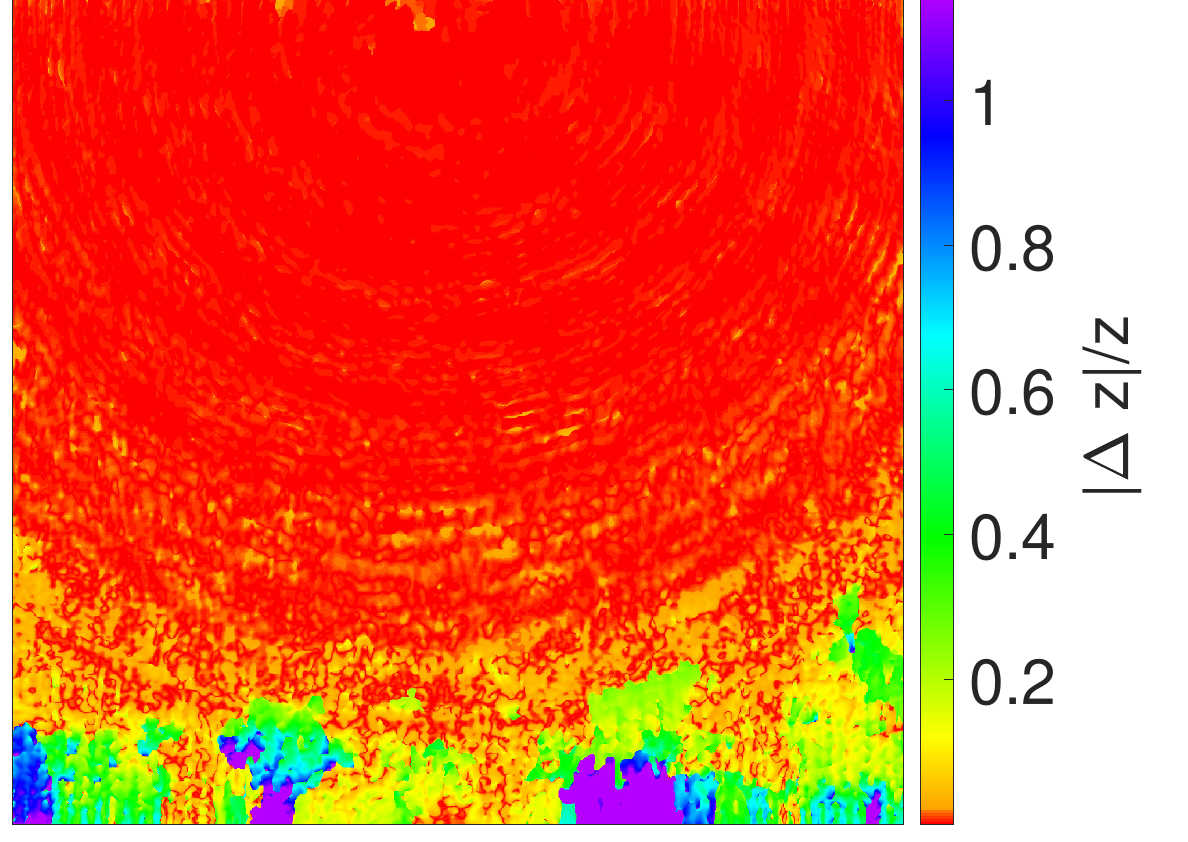}
\label{fig:errors2}
\end{minipage}
\captionof{figure}{a) Deviation $\Delta z$ between the calculated heights from the red and green  channels;  b) Relative error $|\Delta z|/z$ between red and green  channels.}
\label{fig:errors}
\end{Figure}
In the wetted region where the film thickness is greater than $100\,\text{nm}$, both data sets exhibit very good agreement, with an average difference of $\overline{\Delta z}\approx 19.8\,\text{nm}$ over all pixels. However, in the region close to the contact line, the deviation is higher, which illustrates the limitations of the Fourier based evaluation method in this region, as discussed above. Figure~\ref{fig:errors}b shows the relative height error, which indicates that the measurement error is of the same order of magnitude as the film thickness itself in the outer region of the drop.    \\
In the present study, variations in reproducing identical droplet volumes is the limiting factor compared to uncertainties in the evaluation process itself. Therefore, we have refrained from analysing the repeatability, as this does not allow any conclusions to be drawn about the repeatability of the measurement methodology itself.\\
For the chosen droplet volume, it was possible to measure the entire shape of the droplet from the contact line region to its apex within the time interval  $26\,\text{s}-104\,\text{s}$ after deposition. Figure~\ref{fig:shapeSpreading} shows the drop profiles for three chosen spreading times. The data was evaluated from the droplet center point in radial direction, as indicated by  the yellow line that is shown in the inset of Fig.~\ref{fig:shapeSpreading}. No smoothing has been applied here. The curves show the calculated values along a single pixel row.  \\
\begin{Figure}
	\centering
  \includegraphics[width=0.9\textwidth]{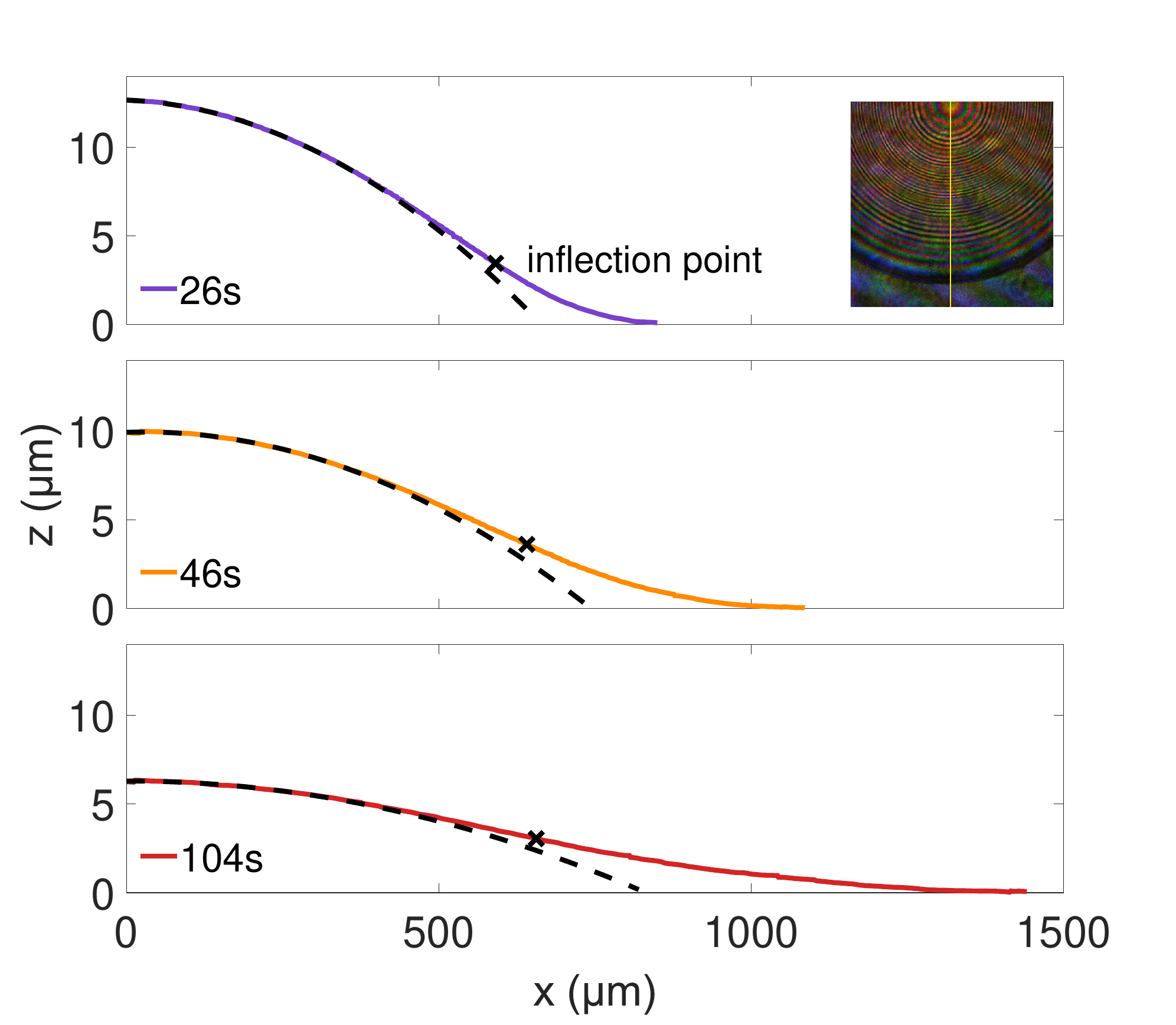}
	\captionof{figure}{Measured droplet profiles for different times after deposition, evaluated along the yellow line shown in the inset picture. The entire droplet shape from the contact line region to the apex was captured. Data points exist for film thicknesses down to $100\,\text{nm}$. As is expected for perfectly wetting liquids, an inflection point exists in all profiles. In the macroscopic region close to the apex, the data is  described well by a spherical cap for all times.}
	\label{fig:shapeSpreading}
\end{Figure}
After just 26 seconds the silicon oil droplet has already spread significantly, leaving a thin film with a height of only a few micrometers. After about two minutes, the radius of the droplet became larger than the field of view,  no longer allowing the entire droplet to be measured. Since the inertial spreading regime of the droplet is expected to be much shorter than one second \cite{PhysRevE.69.016301}, we assume that the exact deposition procedure of the droplet, i.e. the influence of initial velocities, can be neglected over the observed spreading time interval.   \\
The macroscopic part of the droplet in the region close to the apex is well described by a spherical cap, as no significant gravitational deformation of the droplet shape occurs in the present case. The ratio between gravitational and capillary forces is given by the Bond number
\begin{equation}
   \mathrm{Bo} = \frac{\rho g H R}{2\gamma_L},
\label{eq:Bond} 
\end{equation}
where $H$ is the height of the droplet and $R$ the radius of curvature, which is obtained by the spherical cap fit. For $t=104\,\text{s}$, this leads to a Bond number of Bo $\approx 0.077$, which means that capillary forces are dominant and the circular shape is expected close to the apex. However, closer to the contact line region, a strong deviation from the circular shape can be observed, which appears to increase further as the droplet height decreases. \\
The reason for this is that a perfectly wetting liquid drop forms a precursor film ($h<100\,nm$) during spreading. 
The existence of such a precursor film induces an inflection point in the drop profile, as can be seen in Fig.~\ref{fig:shapeSpreading}. Using Eq.~\eqref{hamaker} and the corresponding refractive indices of air, silicone oil and sapphire, we can calculate the microscopic length scale $a\approx 0.83\,\text{nm}$. Thus, using our experimental method only a part of the precursor film can be resolved within the mesoscopic region due to the limitation in film thickness resolution. \\
Usually, the characteristic wetting velocity $U$ is directly calculated from the speed at which the contact line is advancing. However, in a situation where a precursor film is present, this definition is difficult to apply. In Fig.~\ref{fig:times} the spreading radius $R(t)$ at the lowest measurable height of $h=100\,\text{nm}$ is shown. The data  follows closely the power law $R(t) \propto t^{\,0.346}$, which is much faster than the expected theoretical description given by Tanner's law $R(t) \propto t^{\,0.1}$ for the situation where the droplet shape can be approximated by a spherical cap at all times \cite{starov2007wetting}. The dashed curve shows the displacement of the intersection point between the spherical cap curve fit and the x-axis ($z=0$). Also the intersection point, that demarcates  the region of the droplet that can be described by the spherical cap,  spreads faster than expected by Tanner's law. This might occur due to the fact that for such small droplets, the spreading is associated with a considerable height loss of the entire droplet, such that attractive interfacial forces increasingly enhance the spreading process.\\
\begin{Figure}
	\centering
  \includegraphics[width=0.9\textwidth]{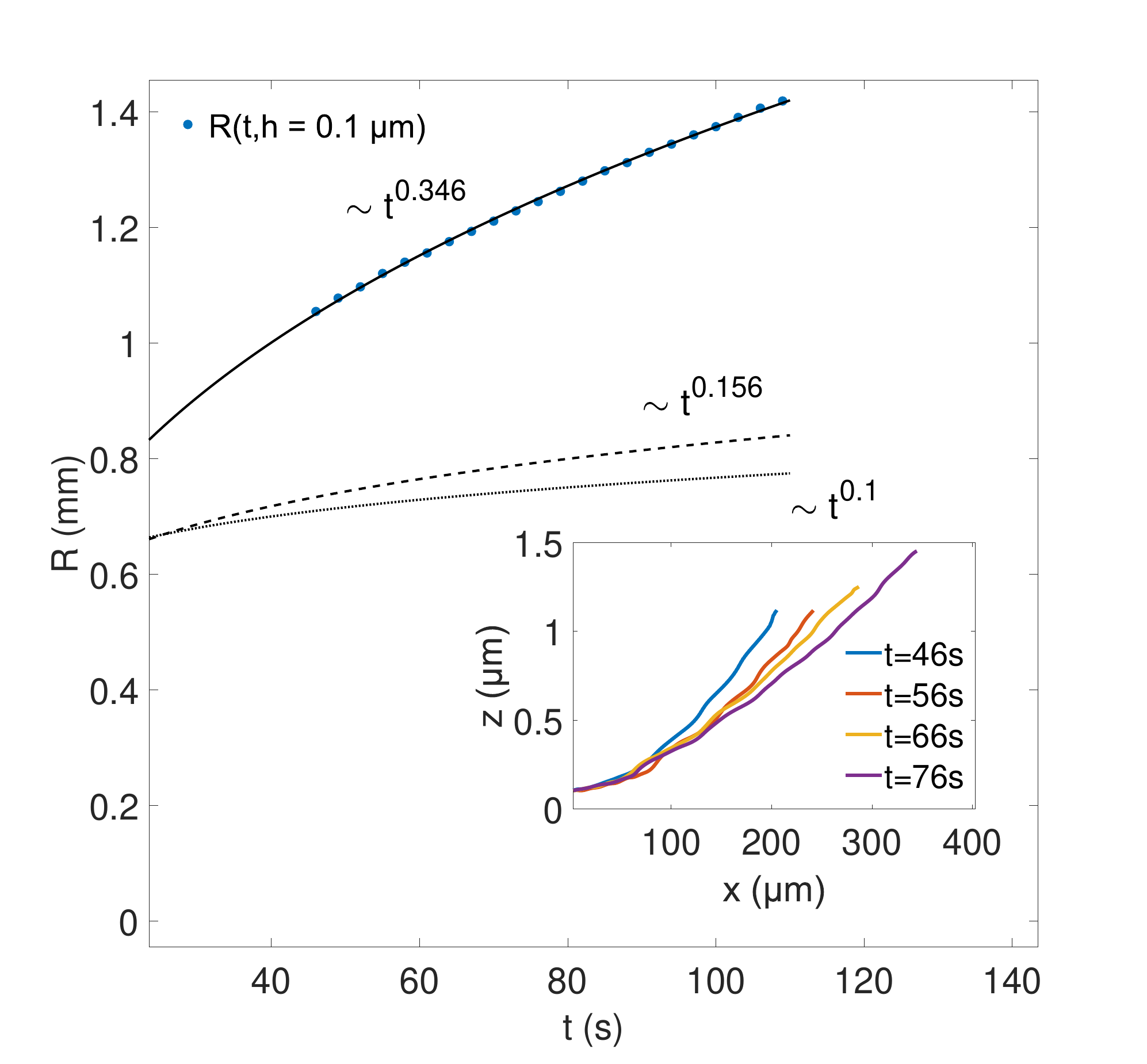}
	\captionof{figure}{Time dependence of the droplet radius evaluated at $h=100\,\text{nm}$ (blue dots). The data points can be described by a power law. The long dashed line corresponds to the intersection point of the spherical cap fit and the substrate ($h=0$). The short dashed line corresponds to the power law $t^{\,0.1}$, which represents the expected value from theory. The inset diagram shows the shape of the liquid front close to the contact line region for different times in the mesoscopic region. The data was shifted in the x-direction to allow for a comparison.}
	\label{fig:times}
\end{Figure}

The inset in Fig.~\ref{fig:times} shows the shape of the liquid front in the mesoscopic region close to the contact line for four different times and therefore for four different spreading velocities. The data is evaluated along a 30 pixel wide region following the yellow centerline, as was shown in the inset of Fig~\ref{fig:shapeSpreading}. To derive the profiles, a one-dimensional Fourier analysis is performed and  the data is processed as described in section~\ref{subsec:lookup}  for each pixel.\\ 
Overall, the present optical technique provides a new opportunity to study  different models of perfect wetting in great detail, because the macroscopic part of the droplet and the microscopic region beyond the inflection point can be observed simultaneously and for any chosen capillary number.

\section{Summary and Discussion}
In this study we have demonstrated a colour interferometric method that utilizes either two or three different wavelengths to study the entire shape of a spreading droplet  from its contact line region ($h>100\,\text{nm}$) to its apex. A Fourier data analysis was used to evaluate fringe images. An average contour deviation of approx. $ 19.8\,\text{nm}$  between the red and green channels could be realized down to a film thickness of $100\,\text{nm}$.
For the wavelengths used in our setup, this allows  direct measurement of a liquid film thickness up to about $1.14\,\mu\text{m}$  without the need for a reference point, where the film thickness is known. The colours were chosen to fit the RGB channels of a colour CMOS camera, such that a unique phase difference could be attained over an optical path length of $3.2\,\mu\text{m}$. It should be noted that this range can be extended by appropriate selection of the wavelengths. 
The use of laser light instead of a broadband light source yields clearly visible interference patterns, even in the case of film heights larger than $10\,\mu\text{m}$, which is possible due to an adequate coherence length of the light source. 
Furthermore, it can be mentioned that the maximum surface slope $h^{\prime}$ that can be measured is related to the spatial distribution of fringes and is therefore proportional to the  magnification. As   $h^{\prime}$ increases, the distance between two adjacent interference fringes decreases in the image plane.\\ Two-colour interferometry has been demonstrated to resolve spatially and temporally the macroscopic and mesoscopic droplet regime during dynamic wetting, providing details of how the free surface of the drop evolves in time. Thus, it is a  powerful technique to study complex dynamic wetting and dewetting processes. This  captures both the macroscopic region that is dominated by capillary forces as well as the mesoscopic region at which interfacial forces prevail. \\
The unique feature of small drops with perfectly wetting properties is that the curvature of the spherical cap and the mesoscopic region are of the same order of magnitude. This is clearly different from the classical case of large droplets, where a vanishing curvature $h''(y\rightarrow \infty)=0$ can be assumed when moving from the mesoscopic towards the macroscopic region \cite{de1985wetting}. We anticipate that it is this change in sign of the interface curvature that leads to an enhanced spreading of a very small drop compared to larger droplets that have a negligible cap curvature. The reasoning behind this is that the switch from a positive to a negative drop curvature leads to an increased Laplace pressure gradient compared to large drops of negligible cap curvature. Therefore, we assume that the rise in pressure gradient inside the drop enhances the spreading velocity of small droplets compared to large drops. This could explain why  the spherical cap for small drops appears to evolve faster  than predicted by Tanner's law. Since the classical picture of an advancing contact line with a constant shape of negligible curvature in the macroscopic region does not hold in the present case, there is a need for extended models for small drops taking into account all curvature effects. \\

\section{Conclusion}
In the present work, we have established and validated a interferometric technique that utilizes multiple wavelengths to study the dynamic wetting dynamics of perfectly wetting liquids, aiming to derive a deeper understanding of the mesoscopic droplet contact line regime, which links the precursor film and the macroscopic bulk liquid of a droplet. \\
Although the  substrate in the present work was  smooth and homogeneous, the image based  method also allows measurements on complex surfaces such as unsteady step-like geometries and different types of rough or coated surfaces, since for small film thicknesses the reconstruction of  film height is possible without any reference points where the film thickness is known. \\
For future work, further investigations on the mesoscopic contact line dynamics will be performed to combine existing macroscopic dynamic contact angle models with microscopic theories.

\section*{Declarations}

\subsection*{Ethical approval}
Not applicable.

\subsection*{Conflict of interest}
The authors declare no competing interests.

\subsection*{Acknowledgments}
The authors gratefully acknowledge financial support from the Deutsche Forschungsgemeinschaft (DFG, German Research Foundation) within the Collaborative Research Centre 1194 – Project-ID 265191195 “Interaction of Transport and Wetting Processes”.\\

\subsection*{Availability of data and materials}
Data will be made available upon request to the corresponding author.

\subsection*{Author contributions}
TR performed the experiments and took the lead in data analysis, interpretation and writing the manuscript. 
MF supported the data interpretation and reviewed the manuscript. 
PS contributed to the interpretation of results.
CT supported the data interpretation and reviewed the manuscript.
JH conceived the project, supervised the conceptualization of this study, the data interpretation and reviewed the manuscript.

\printbibliography

\end{multicols}
\end{document}